# SAGNAC INTERFEROMETER ON THE HOLOGRAPHIC BRAGG GRATING

E. A. Tikhonov, A. K. Lyamets, Institute of physics National Academy Sci., Ukraine, Kiev, <etikh@live.ru>

**SUMMARY** – The ring interferometer with zero optical path difference known as Sagnac one (INS) is offered with a diffraction splitting of the entering light beam. As the beamsplitter a transmission holographic Bragg grating is used. Conditions of normal operation of INS achieve under equal intensity of beam copies and the adjustable phase shift between them in its two interferometer shoulders. These conditions are met with the holographic grating, which provides the phase shift $\pi$ on the central Bragg wavelength. Experimental approbation of the modified INS validates the expected results.

**INTRODUCTION AND MOTIVATION OF WORK**

Thick holographic gratings with Bragg's diffraction can be used for obtaining coherent copies of monochromatic various directed beams with constant total but adjustable individual power and phase shift between them. Characteristics of phase holographic gratings are described by the theory of the coupled waves in Kogelnik's model [1]. Experimental check of the Kogelnik's model applicability in relation to the photopolymer holographic materials and elements, recorded in real time, has been provided in work [2]. On this basis and practical opportunities of authors to record various holographic gratings some optical schemes of different functional destination have been offered [3-9].

In this work operation of Sagnac interferometer using the transmission holographic Bragg phase grating (HBG) as diffraction beam splitter is studied [10]. The scheme of the INS is presented on fig. 1.

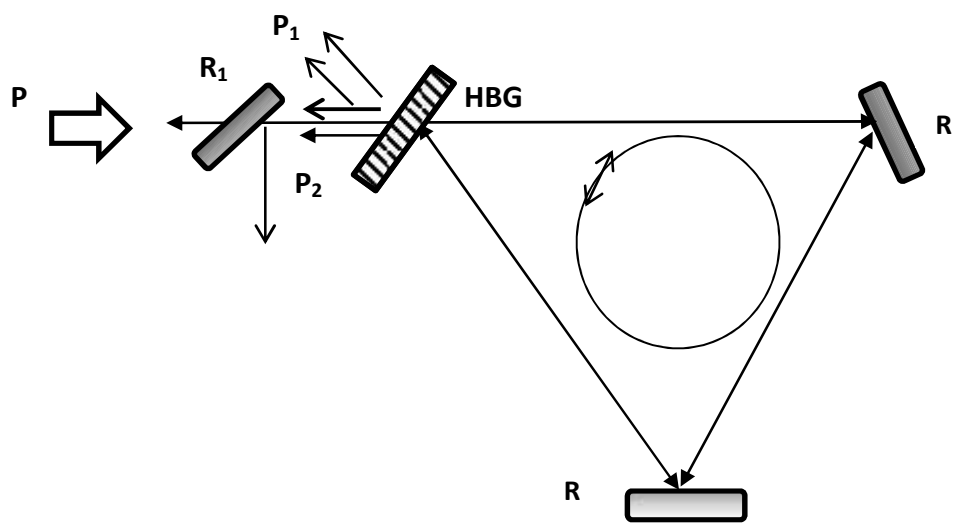

Fig. 1. The optical scheme of INS with HBG: R-reflective dielectric mirrors,
$P_2$ and $P_1$ – operational and control outputs of radiation from INS, P- input beam, HBG as beam splitter, $R_1$ - mirror for check of output power $P_2$

HBG with harmonic index modulation is characterized by the following dependence of the diffraction efficiency (DE) as function of phase shift /1/:

$$\eta_{s,p} = (1 + \xi^2/\Phi_{s,p}^2)^{-1} \sin^2[\Phi_{s,p}(1 + \xi^2/\Phi_{s,p}^2)^{0.5}] \quad (1)$$

where $\xi = 2\pi n \Delta T \sin\theta / \lambda$, $\Delta\theta = \theta^* - \theta$, - generalized and direct angular Bragg's detuning, $\nu_{s,p}$ – phase foray in the lack of detuning respectively for s and p of polarisation, , T- grating thickness, n and λ- material index of refraction and wavelength of measurements.

In lack of detuning (ξ=0) power of the diffracting beam is determined by the magnitude of phase foray $\nu_{of\ s,\ p}$. If HBG on given wavelength provides phase foray π/2, DE = 1 and in INS there is only one diffracted beam. For normal work of INS it is required 2 equal about intensity and polarization beams with identical and opposite phases at its two exits. Therefore to split the input beam into 2 almost power equal it agrees (1) at angular detuning of INS ξ≅π/4. At similar setup and tuning of GBR at output beam $P_1$ includes the couple of beams equal on power with phase shift between them π/2 which provides only partial damping of beams unlike classical INS with beam splitter on transmission dielectric mirror; at the output beam $P_2$ consist the similar couple of beams, but with zero phase shift between. Thus, at setup of typical HBG for work INS the necessary conditions about equality of power is satisfied, but condition about the antiphase shift π is not reached.

If the phase difference at single diffraction on HBG was equal π/2, the total phase difference after second diffraction of beam output $P_1$ would reach π, and in the direction would be the complete interferential damping of $P_1$ power with the complete recovery of power at $P_2$ exit where in couple of beams the zero phase difference remains always. Because for typical HBG DE=100% is reached under specified value of phase foray π/2 (fig. 2.), the second beam in zero diffraction order does not arise practically.

However the mild implementation of INS requirements about output power damping $P_1$ under decrease of interferential contrast is perhaps. Let's give simple estimates of similar compromise, using the accepted determination of interferential contrast through intensity and phase shift between beams:

$$K = 2\sqrt{I_1 I_2} \cos\varphi / (I_1 + I_2) \quad (2)$$

Let's express values of beam intensity through the diffraction efficiency DE = $\eta_{s,\ p}$:

$$K = 2\sqrt{\eta_{s,p}(1-\eta_{s,p})} \cos\varphi \quad (3)$$

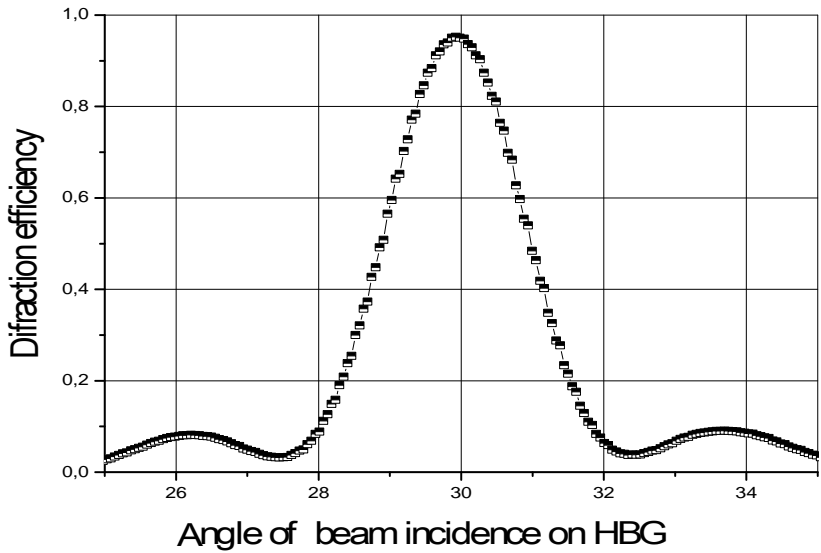

Fig. 2. Angular DE dependence of phase HBG as (632,8nm, s-polarization).

After substitution (1) and simple transforms we receive:

$$= \frac{2}{a(\prime)} \sin^2(2\epsilon_{s,p} a(\varsigma)) \cos_{\pi} \qquad (4)$$

In (4) for brevity designation is used here $a(\varsigma) = (1 + \varsigma^2/\epsilon^2_{s,p})^{0.5}$.

In the considered ring INS phase shift provides the new optical element – the transmision phase grating (in comparison with classical INS) therefore the phase angle of interferential summation $\theta$ can be defined through the phase foray of HBG $_{\pi} \approx 2\epsilon_{s,p} a(\varsigma)$. Finally we receive the following expression for contrast in INS with HBG:

$$= \sin(4\epsilon_{s,p} a(\varsigma)) \sin(2\epsilon_{s,p} a(\varsigma)) \qquad (5)$$

From (5) follows that contrast ≈1 in our INS based on HBG is achievable under following condition:
$$\epsilon_{s,p} a(\varsigma) \approx (2m+1)f/8, \quad m=0,1,2... \qquad (6)$$

The condition (6) specifies that the angular provision of maximal contrast can be aside DE maximum where splitting of incident beam into copies with unequal amplitudes takes place that reduces contrast relation of INS.

Other opportunity to reach 100% of contrast in INS with HBG consists in use of grating with phase foray at center Bragg wavelength $v_{s,p} > \pi/2$, is possible owing to so-called "overmodulation" or simple increase in thickness of grating /2/. In fig. 3a. it is shown calculated and experimental options of the discussed HBG.

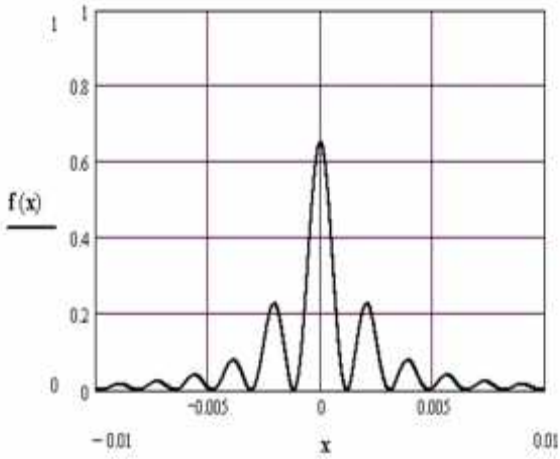 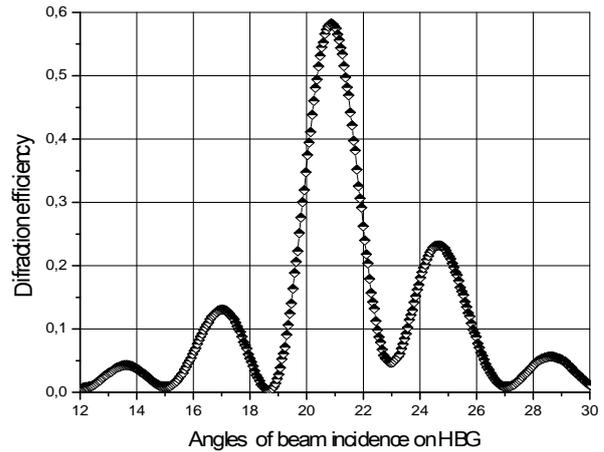

Fig. 3a. Calculated and experimental options DE of HBG with phase foray 2,2 rad (>π/2) (s-polarization, 632,8nm)

It is possible to build INS version on HBG having phase jump π at center Bragg frequency. The calculated contour of angular dependence of DE for such grating is shown in fig. 4. AS can be seen the similar grating at angular tuning on the every from two symmetric maxima provides at the same time strict splitting of beam into two equal power and with antiphase shift between them at P1 exit (as single phase shift in these positions is π/2 or 3π/2 that provides constructive interference at $P_2$ INS exit the and total destructive interference at $P_1$ exit.

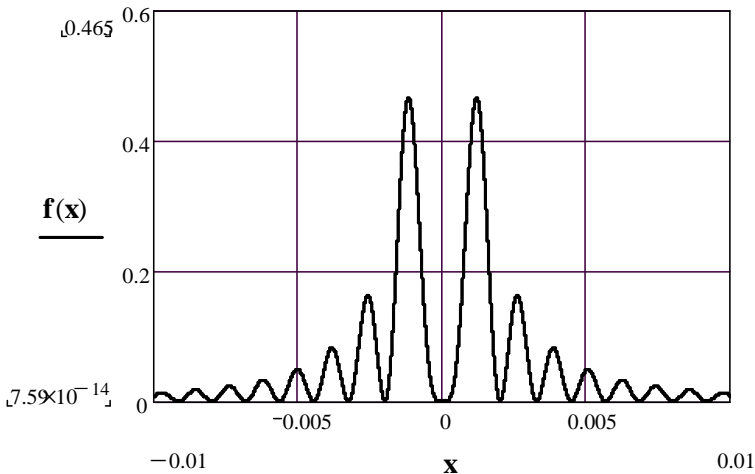

Fig. 4. Calculated contour of DE for HBG with phase jump =π radian at center Bragg frequency

**EXPERIMENTAL TESTING**

A change of output power at tuning turn of GBR in the vicinity of Bragg angle was registered on main ($P_2$) and control ($P_1$) INS exits (fig. 5.). Registration of radiant power was carried out by silicon photo diodes FD-24 in the mode of reverse current with ADT and the computer (linearity on light flux response was about 3 orders of magnitude). Processing of results was carried out by the Origin 8.0 program. Received results are presented below on fig. 5-7. For the case INS with GBR in lack of detuning ($\xi=0$) the typical HBG (fig. 2.) provides the phase difference between beam copies $\pi/2$ and all power concentrated on $P_1$ exit. At symmetric angular detuning around the main maximum of DE this HBG at the exits of INS two couples of output power appear in INS (A and B) with antiphase nature of change of power in main $P_2$ and control $P_1$ shoulders. Their angular positions and magnitudes are defined by ratio of amplitudes and phases between the corresponding couples of beams. The angular location of these extrema in this case is at $\approx 29^0$ and $\approx 31^0$ and corresponds to phase foray due to diffraction $\pi/4$ and $(3/4)\pi$, providing also splitting of input beam power approximately in half. Far from main Bragg resonance at incidence angles $\leq 25^0$ or the $\geq 34^0$ splitting of the incident beam is rather small due to quite low DE, and, respectively, there is no formation of couples of equal beams at both exits of INS. The same is true for small registered minimax of INS on the angle $26,5^0$ and $33^0$ in spite of that phase shift between copies of beam is optimal $\pi/2$ but power of every one is rather different. Thus at the control exit of INS $P_1$ the incomplete destructive interference of beam power is observed due to insufficient difference in phase equal $\pi/2$. At the main exit of INS $P_2$ both beams stay in phase (both have experienced phase shift $\pi/4$), however some part of input power are lost at $P_1$ exit owing to an incomplete interferential suppression.

Minimaxes at the exits of ring INS are possible only in situation which is regulated by condition (5). For this HBG it is achievable partially as at setup DE=50 % when phase shift directs to $\pi/4$ (or $3\pi/4$) (at the left and on the right relative to Bragg angle. In this case at twice diffraction of one of beams full phase difference is $\pi/2$ that is not enough to total suppression. In reality minima of suppression of output power $P_1$ are observed at $29^0$ and $31^0$. Meanwhile left minimum $P_1$ at $29^0$ is approximately 0,07v, right minimum at $31^0$ is only 0,31v and approximately equal to $P_2$ maximum at the control exit, respectively contrasts are equal 86% and 17% . In other words, effective suppression of beam power at the main exit of INS does not happen with a such HBG though noticeable interferential damping about 86% in the left wing is observed.

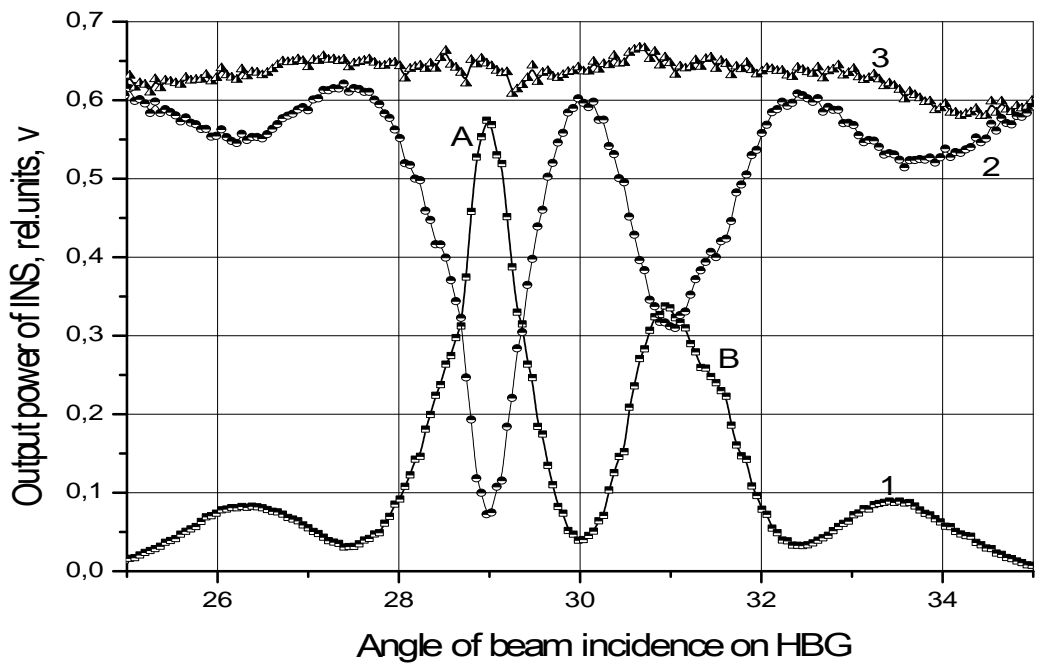

Fig. 5. Interferential summation and subtraction of power in INS shoulders: 1 – output power of $P_2$, 2- output power $P_1$ according to fig. 1., 3-total power of P1 and P2. Measurements are made on wavelength 632,8nm the s-polarized radiation of the He-Ne laser. The contrast relation on anglesl of $29^0$ - 86% and $31^0$ - 17%

On the right wing at $31^0$ low level of beam suppression it is possibly obliged to uncontrollable phase shift owing to low plane-parallel HBG (two substrates and layer of polymeric grating 50-100mkm).

At interferential tuning of INS the contrast relation for A and B extrema was interchanged the angular positions. Thin interferential structure in the region of extremum A and B fig. 6. arose at careful interferential tuning of INS. The contrast relation for the brightest extremum remained invariable.

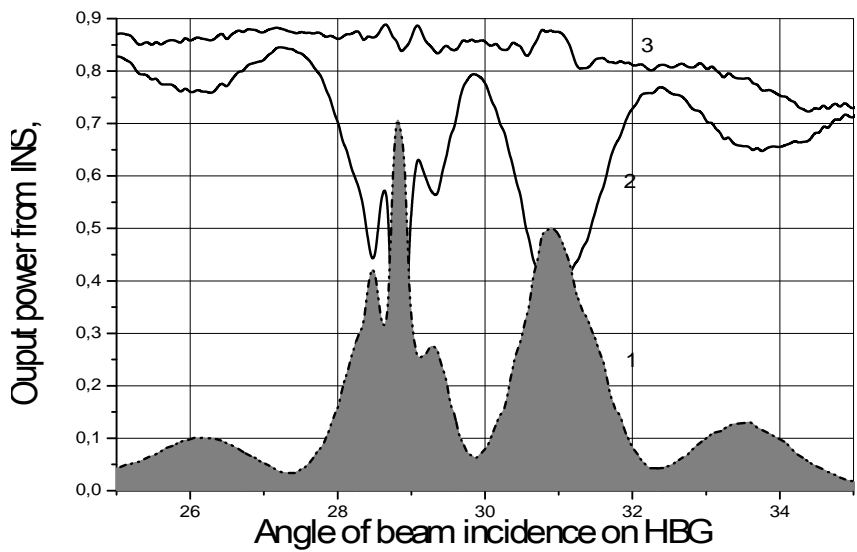

Fig. 6. Interferential structure of antiphase behavior in INS shoulders. (1 P2, 2 P1, 3 – P1 + P2) Other conditions were similar for experiences in fig. 5.

Considerably higher contrast relation for output power result in INS shoulders has been reached with HBG (fig. 3b) which is recorded with noticeable "overmodulation" so phase shift in maximum of DE is ≈2,2рад=0,7π. Setup of INS is made on detuning $\xi$ at which copies of initial beam become equal on power. With this grating the contrast relation of power at the exit of INS reached ≈98,8% and remained within ≈0,5⁰.

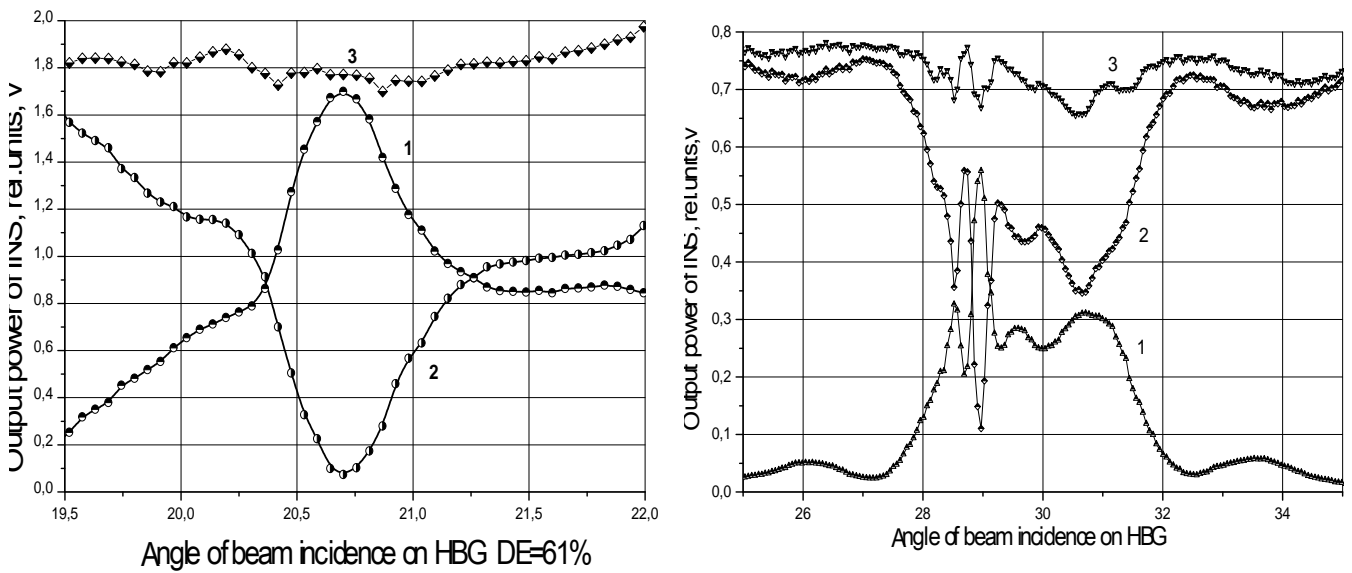

fig. 7a, b) Output power in main exit $P_2$ and control exit $P_1$ of INS (1 and 2) with HBG having DE=61% at phase shift on Bragg angle of 0,7π. (s-polarization, 632,8nm) 3-total power of P1+P2) b) Interferential effects of summation and subtraction of powers in shoulders of $P_2$ and $P_1$ of INS.

Deficiency of HBG with phase shift equal π at center Bragg frequency in the made experiments has not allowed to show the brighter result of interferential suppression of beam couple at the control exit this modification of INS. Let's notice that in this INS, as well as in its classical option, suppression takes place in the direction of the transmission of INS and reflection amplification – in backward direction to the input beam.

**FINAL NOTES**

The offered INS option with diffraction beam splitter of light on the basis of phase transmission holographic grating is represented perspective system as for the traditional applications IN mentioned partially among the quoted works /9,10/, so for other applications possible owing to new properties of this system. The offered INS unlike classical analog, can bring the regulated amplitude difference and the phase shift between beam proving in change of contrast of interference. For example, entering of non-mutual elements into ring of the interferometer one can be welcomed to change of the contrast relation which is possible to compensate reasonably turn of phase grating splitter and to estimate thus quantitatively similar effect. It is clear that experimental implementation of such INS assumes use of the goniometer of minute and higher angular resolution power.

**THE QUOTED LITERATURE**